\documentclass[ApJ]{emulateapj}

\newcommand{\crhydro}{{\it CR-hydro-NEI}}
\newcommand{\chandra}{{\it Chandra}}
\newcommand{\RXJ}{RX~J1713.7$-$3946}

\newcommand{\synch}{synchrotron}
\newcommand{\pion}{pion-decay}
\newcommand{\IC}{inverse-Compton}
\newcommand{\gamray}{$\gamma$-ray}

\newcommand{\SC}{self-consistent}

\newcommand{\alf}{Alfv\'en}
\newcommand{\pmax}{p_\mathrm{max}}
\newcommand{\Kep}{K_\mathrm{ep}}

\def\lsim{\raise0.3ex
  \hbox{$<$\kern-0.75em\raise-1.1ex\hbox{$\sim$}}\,}

\def\gsim{\raise0.3ex
  \hbox{$>$\kern-0.75em\raise-1.1ex\hbox{$\sim$}}\,}

\usepackage{ulem,amssymb} 
\usepackage{color} 

\slugcomment{Accepted for Publication in The Astrophysical Journal}

\shorttitle{\crhydro\ Model of Tycho}
\shortauthors{Slane et al.}

\begin{document}

\title{A CR-hydro-NEI Model of the Structure and Broadband Emission from
Tycho's SNR} 

\author{P.~Slane \altaffilmark{1},
S.-H.~Lee\altaffilmark{2},
D.~C.~Ellison\altaffilmark{3},
D.~J.~Patnaude \altaffilmark{1},
J.~P.~Hughes,\altaffilmark{4},
K.~A.~Eriksen\altaffilmark{4,5},
D.~Castro\altaffilmark{6},
and S.~Nagataki\altaffilmark{2}
}

\altaffiltext{1}{Harvard-Smithsonian Center for Astrophysics, 
60 Garden Street, Cambridge, MA 02138, USA;
slane@cfa.harvard.edu; dpatnaude@cfa.harvard.edu}

\altaffiltext{2}{RIKEN, Astrophysical Big Bang Laboratory, 2-1
Hirosawa, Wako, Saitama 351-0198, Japan; shiu-hang.lee@riken.jp;
shigehiro.nagataki@riken.jp}

\altaffiltext{3}{Physics Department, North Carolina State
University, Box 8202, Raleigh, NC 27695, USA;
don\_ellison@ncsu.edu}

\altaffiltext{4}{Department of Physics and Astronomy, Rutgers University,
Piscataway, NJ 08854-8019, USA; jph@physics.rutgers.edu}

\altaffiltext{5}{XTD-IDA, Los Alamos National Laboratory, P.O. Box 1663, Los
Alamos, NM 87545, USA; keriksen@lanl.gov}

\altaffiltext{6}{MIT-Kavli Center for Astrophysics and Space Research, 77
Massachusetts Avenue, Cambridge, MA 02139, USA; castro@mit.edu}

\begin{abstract}
Tycho's supernova remnant (SNR) is well-established as a source of
particle acceleration to very high energies. Constraints from
numerous studies indicate that the observed $\gamma$-ray emission
results primarily from hadronic processes, providing direct evidence
of highly relativistic ions that have been accelerated by the SNR.
Here we present an investigation of the dynamical and spectral
evolution of Tycho's SNR by carrying out hydrodynamical simulations
that include diffusive shock acceleration of particles in the
amplified magnetic field at the forward shock of the SNR.  Our
simulations provide a consistent view of the shock positions, the
nonthermal emission, the thermal X-ray emission from the forward
shock, and the brightness profiles of the radio and X-ray emission.
We compare these with the observed properties of Tycho to determine
the density of the ambient material, the particle acceleration
efficiency and maximum energy, the accelerated electron-to-proton
ratio, and the properties of the shocked gas downstream of the
expanding SNR shell.  We find that evolution of a typical Type Ia
supernova in a low ambient density ($n_0 \sim 0.3 {\rm\ cm}^{-3}$),
with an upstream magnetic field of $\sim 5\ \mu$G, and with $\sim
16$\% of the SNR kinetic energy being converted into relativistic
electrons and ions through diffusive shock acceleration, reproduces
the observed properties of Tycho.  Under such a scenario, the bulk
of observed $\gamma$-ray emission at high energies is produced by
$\pi^0$-decay resulting from the collisions of energetic hadrons,
while inverse-Compton emission is significant at lower energies,
comprising roughly half of the flux between 1 and 10 GeV.

\end{abstract}


\section{Introduction}
Efficient acceleration of charged particles in the shocks of supernova
remnants (SNRs) has long been cited as a likely process through
which a significant fraction of Galactic cosmic-rays (CRs) are
accelerated.  The evidence for such energetic particles is compelling.
Radio emission from SNRs originates from electrons with energies
$E_e > 1$~GeV, while observations of nonthermal X-ray emission from
SNRs reveal electrons with energies exceeding tens of TeV.  Evidence
for energetic ions accelerated in SNRs is more elusive because of
the low radiation efficiency for such particles.  The observed GeV
and TeV $\gamma$-ray emission from many SNRs -- particularly those
known to be located in dense environments -- is consistent with the
presence of energetic protons that produce $\gamma$-rays through
the decay of neutral pions created in collisions with ambient nuclei.
But $\gamma$-rays can also be produced from the energetic electron
population, through inverse-Compton (IC) scattering of ambient
photons or through nonthermal bremsstrahlung. Modeling of the
broadband spectra from such SNRs, in order to ascertain the nature
of the $\gamma$-ray emission, is complicated and has led to mixed
interpretations, making the evidence for ion acceleration controversial
in some cases. Gamma-ray emission from some SNRs known to be
interacting with molecular clouds seems to require a significant
component from pion decay \citep[e.g.][]{AbdoEtalW51C2009, cs10, AbdoEtalCasA2010,
GiulianiEtal2011, Ackermann_Science},
with some remnants requiring nearly all of the SNR kinetic
energy to be converted to relativistic electrons for IC emission
to dominate the flux \citep{CSCF2013,ASC2013}.
For W44 and IC~443, the $\gamma$-ray spectra show clear evidence
of a kinematic ``pion bump,'' firmly establishing the presence of
energetic ions in these remnants \citep{Ackermann_Science}.

\begin{figure*}[t]
\centering
\setlength\fboxsep{0pt}
\setlength\fboxrule{0.0pt}
\fbox{\includegraphics[width=6in]{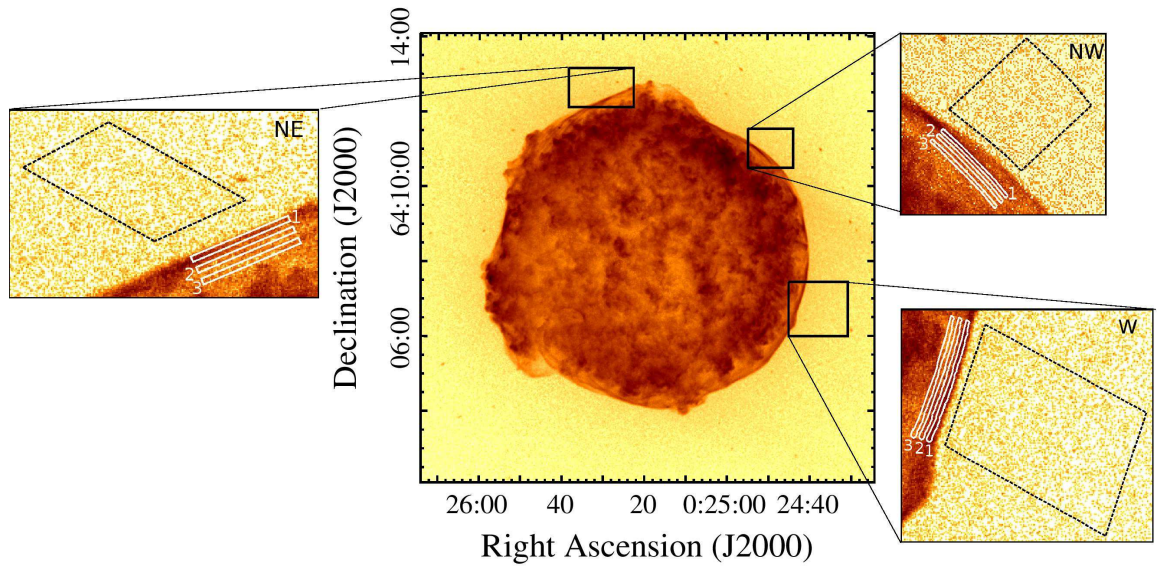}}
\caption{
\chandra\ image of Tycho's SNR. Insets identify regions used for
spectral extraction, with dashed regions indicating background.
}
\label{fig1}
\end{figure*}

Ion acceleration can have observable dynamical effects on SNR
evolution, since this process results in less thermal heating of
the swept-up gas and an increased compression ratio in the postshock
region. X-ray studies of 1E0102 \citep{HRD2000} reveal electron
temperatures that are much lower than expected from the observed
expansion velocities of the remnants, for example, indicating that
a large fraction of the shock energy has gone into something other
than thermal heating of the gas. In addition, as we discuss in more
detail below, X-ray studies of Tycho's SNR \citep{WarrenEtal2005}
demonstrate that the ratio of the forward shock (FS) radius to that
of the contact discontinuity (CD), as well as that of the reverse
shock (RS), is smaller than expected, consistent with results
expected from efficient ion acceleration \citep{DEB2000, EDB2004}.
Finally, deep \chandra\ observations of Tycho reveal a complex of
regularly-spaced stripe-like nonthermal structures in the projected
interior of the remnant \citep{EriksenEtal2011}.  The spacing of
these stripes may correspond to the gyroradii of $10^{14} - 10^{15}$~eV
protons in an amplified magnetic field (Eriksen et al. 2011),
although \citet{beopu2011} suggest that the structures may be the
result of anisotropic magnetic turbulence produced by instabilities
driven by CR protons with energies of $\sim 10^{15}$~eV.  The maximum
energy to which such ions are accelerated in a particular SNR is
of critical importance to our understanding of the role SNRs play
in producing ions with energies approaching the knee of the CR
spectrum.  Combined with modeling of the broadband emission that
appears to imply that pion-decay dominates the $\gamma$-ray flux
\citep{MC2012, GiordanoEtal2012, BKV2013, ZhangEtal2013}, these
observations have thus led to the conclusion that Tycho's SNR is a
particularly important testbed for models of particle acceleration
in SNRs.

The thermal emission from material compressed by the FS provides
particularly important information on particle acceleration in SNRs.
Because the postshock temperature is reduced in the case of efficient
acceleration, the emission characteristics of the plasma are modified.
Moreover, the ionization state of the gas is modified by both the
reduced temperature and the higher density associated with the
increased compression ratio \citep{PES2009}. As a result, self-consistent
treatment of the thermal emission is crucial in any effort to model
the effects of cosmic ray acceleration in SNRs.  This is of particular
importance in assessing the nature of any observed $\gamma$-ray
emission because of the critical dependence on density shared by
both $\pi^0$-decay emission and thermal X-ray emission. For example,
the lack of observed thermal X-ray emission from \RXJ\ eliminates
$\pi^0$-decay as a significant contributor to the $\gamma$-ray
emission \citep{EPSR2010, ESPB2012} unless the postshock medium is
filled with cold, clumped gas \citep{InoueEtal2012}. Conversely, the
density required to produce the observed thermal X-ray emission
from H-like ions of Si in CTB~109 is sufficiently high for $\pi^0$-decay
to account for roughly half of the observed $\gamma$-ray flux (Castro
et al. 2012). Any complete picture of the cosmic-ray modified
emission and dynamical evolution of Tycho's SNR must include a
self-consistent treatment of thermal X-ray emission.

Tycho's SNR is the product of SN~1572. Based on historical records
of its light curve, the remnant has long been understood to have
resulted from a Type Ia event \citep{R-L2004}, corresponding
to the thermonuclear destruction of a C-O white dwarf star. This
has been confirmed through direct measurements of the supernova
spectrum, taken from light echo measurements \citep{KrauseEtal2008}
that identify it as belonging to the normal Type Ia class of SNe.
As with many Galactic SNRs, the distance to Tycho's SNR is rather
uncertain. Most estimates fall in the $2 - 5$~kpc range with recent
estimates of $4 \pm 1$~kpc based on observed ejecta velocities and
proper motion measurements \citep{HayatoEtal2010}, $3.8^{+1.5}_{-0.9}$~kpc
based on light echo measurements (Krause et al. 2008), and $2.5 -
3$ kpc based on kinematic methods \citep{TL2011}. Morlino \&
Caprioli (2012) estimate $d = 3.3$~kpc based on broadband modeling
of the spectrum (see Section 4 below), although their analysis is
based on the \citet{TM99} parameterization of the SNR
evolution rather than a self-consistent hydrodynamical treatment
that includes the effects of the CR acceleration, such as that used
here.

The X-ray emission from Tycho's SNR (Figure 1) is dominated by
ejecta \citep{HG97, DecourchelleEtal2001, HwangEtal2002}, accompanied
by thin filaments of synchrotron emission from extremely energetic
electrons accelerated at the forward shock \citep{cad04a,
EriksenEtal2011, beopu2011}.  The mean angular radius of the remnant
is $251^{\prime\prime}$, with an azimuthal variation of about $\pm
16^{\prime\prime}$ \citep{WarrenEtal2005}. Comparison of the ejecta
density and composition structure with hydrodynamical models for
the evolution and nucleosynthesis models for the ejecta produced
in different classes of Type Ia explosions indicate that the remnant
is the result of a delayed-detonation explosion \citep{Badenes2006}
with a $\sim 10^{51}{\rm\ erg}$ explosion expanding into an ambient
density $n_0 = 0.85 - 2.1 {\rm\ cm}^{-3}$, although most other
studies indicate lower densities: $n_0 \lsim 0.3 {\rm\ cm}^{-3}$
based on limits to the thermal X-ray emission (Cassam-Chena\"{\i}
et al.  2007); $n_0 \lsim 0.4 {\rm\ cm}^{-3}$ based on the $\gamma$-ray
flux (V\"{o}lk et al. 2008); $n_0 \lsim 0.2 {\rm\ cm}^{-3}$ based
on measurements of the SNR expansion index $m$, where $R \propto
t^m$ \citep{KatsudaEtal2010}; and  $n_0 \sim 0.1 - 0.2 {\rm\
cm}^{-3}$ (except in distinct regions of known dense clump interactions)
based on the ratio of 70 to 24 $\mu$m flux ratios from postshock
dust in Tycho \citep{WilliamsEtal2013}.

Gamma-ray emission from Tycho has been identified by observations
with VERITAS (Acciari et al. 2011) and the Fermi LAT
(Giordano et al. 2012), and models of the broadband spectrum
indicate that the $\gamma$-ray flux is dominated by emission
from $\pi^0$-decay (Morlino \& Caprioli 2012; Berezhko et al. 2013; 
Zhang et al. 2013), although the details of the models used to
reach this conclusion differ considerably. In addition, arguments
have been made for models in which the $\gamma$-rays may be dominated
by IC emission \citep{AD2012}.

Here we present a study of the radial structure, evolution,
and broadband emission from Tycho's SNR using hydrodynamical
simulations described in Section 2. In Section 3 we describe our
modeling approach and summarize the application of these models to
the broadband spectral energy distribution (SED) for Tycho as well
as the X-ray emission from the post-shock region of the SNR blast
wave. We discuss the results of our modeling efforts in Section 4,
in the context of previous studies of Tycho, and our conclusions
are presented in Section 5.  We confirm, in what we believe to be
the most complete, spherically symmetric, broadband model of this
SNR yet performed, that the bulk of the $\gamma$-ray emission is
from hadronic processes with a significant fraction of $\gamma$-ray
emission contributed by leptonic processes at GeV energies.  Other
important properties of the SNR, such as ambient density, magnetic
field strength, the DSA efficiency, and the relativistic
electron-to-proton ratio, are also constrained.

\section{The \crhydro\ Model}

To address the evolution, particle acceleration, and broadband
emission for Tycho's SNR, we have used the \crhydro\ code that models
the SNR hydrodynamics with a version of the VH-1 hydro code
\citep[e.g.,][]{BE2001} modified to include the effects of non-linear
diffusive shock acceleration (DSA) using a semi-analytic solution
based on the treatments from \citep{BGV2005} and \citep{CBAV2009}.
The resulting nonthermal proton and electron spectra,
coupled with the calculated (amplified) magnetic field and assumed
ambient photon fields, are used to calculate the synchrotron,
bremsstrahlung, inverse-Compton, and $\pi^0$-decay emission. The
thermal X-ray emission is calculated by following the ionization
of the shocked gas through the hydro simulation and coupling this
to a non-equilibrium ionization emission code (e.g., Ellison et al.
2007; Patnaude et al. 2009).  A full description of the \crhydro\
code can be found in \citet{LEN2012}. The approach here is similar
to that used for investigations of \RXJ\ (Ellison et al. 2012),
CTB~109 (Castro et al. 2012), and Vela~Jr \citep{LSENP2013}.

\begin{figure*}[t]
\centering
\setlength\fboxsep{0pt}
\setlength\fboxrule{0.0pt}
\fbox{\includegraphics[width=3in]{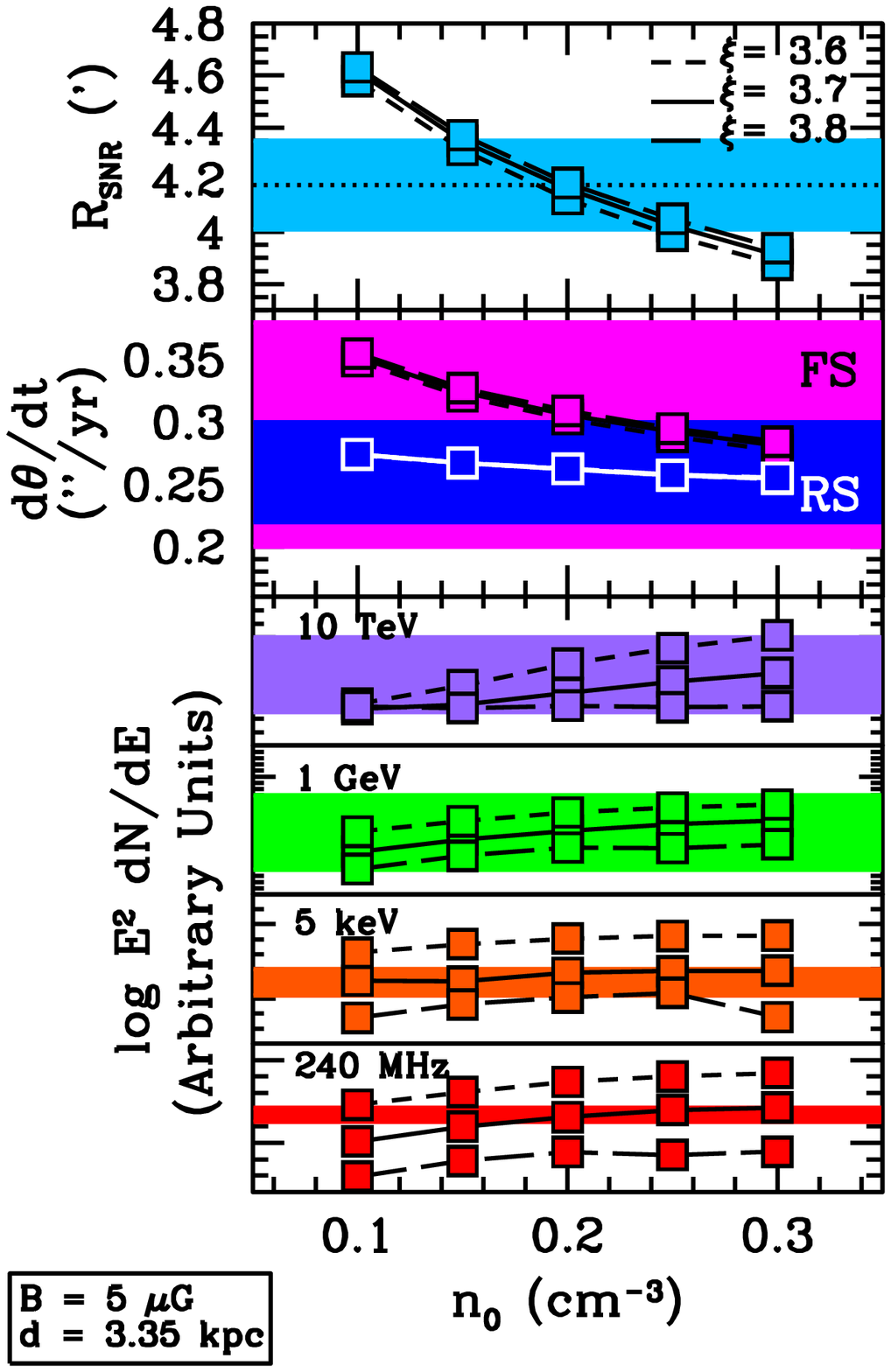}
\hspace{0.4in}
\includegraphics[width=3in]{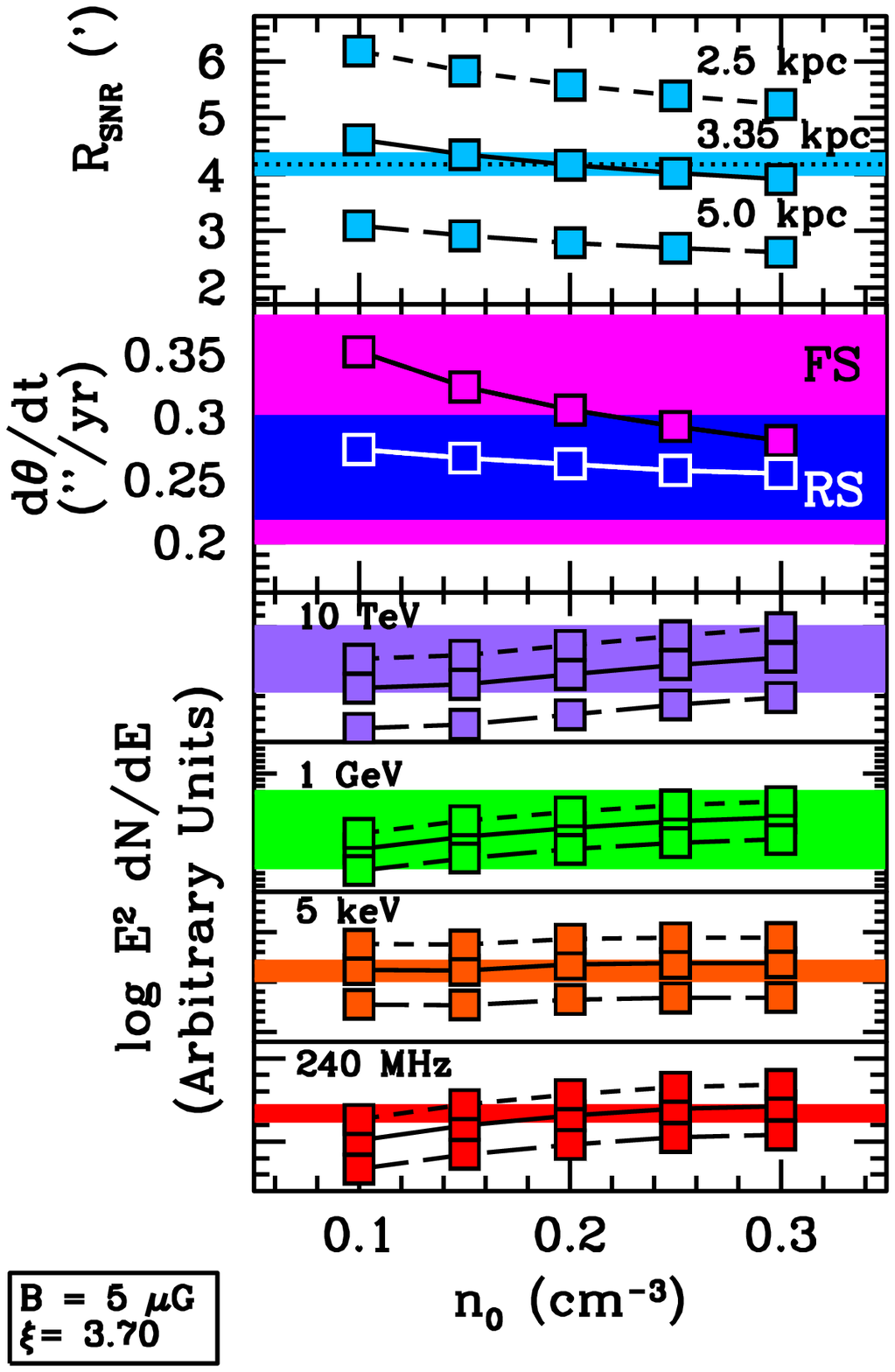}}
\caption{
Comparison of model predictions for the angular size, $R_\mathrm{SNR}$,
the angular expansion rate, $d\theta/dt$, for the FS and RS, and
the radio, X-ray, and $\gamma$-ray fluxes for Tycho's SNR as a
function of ambient density.  The left panel shows model results
for different DSA injection parameters, $\xi$, for fixed values for
the distance and upstream magnetic field. Note that smaller values
of $\xi$ imply larger acceleration efficiencies (see Table 1).  The
right panel shows results for different distance values for fixed
efficiency and magnetic field values.
\label{fig2}}
\end{figure*}

We note that while our model is spherically symmetric, it is
inhomogeneous in radius.  We begin with a radial ejecta density
distribution (assumed here to be exponential, with a total mass of
$1.4 M_\odot$), and all parameters including the density, temperature,
magnetic field, and ionization state, evolve as the remnant ages
and shocked material advects downstream from the forward shock. At
the current age, the relativistic electrons and ions, as well as
the thermal plasma, emit from a ``continuous zone" environment
between the CD and FS.  Rather then assuming a single
power law with an exponential cutoff for the acceleration CRs, as
in some models (e.g. Zhang et al. 2013), with the power law index
and cutoff parameter fixed to match the data, we determine the
evolving full particle spectrum, spatially-resolved and integrated
over time, using a \SC\ model.

\section{Modeling}

The primary parameters for driving the simulations are the SNR
distance ($d$), the upstream (i.e., unshocked) density ($n_0$) and
the upstream magnetic field ($B_0$; both assumed constant here),
the kinetic energy of the explosion ($E_{51}$, in units of
$10^{51}$~erg), and the DSA injection efficiency parameter ($\xi$).
We investigated a range of distances from 2.5 - 5 kpc, and explored
a grid of values for the other key parameters for each assumed
distance until the observed angular radius was reproduced at the
known age of Tycho.  We compared the angular positions of the RS
and CD for each model with the measured values from Warren et al.
(2005), and the predicted angular expansion speed with measurements
from Katsuda et al. (2010), and rejected models for which the
discrepancy was larger than $\sim 0.5$~arcmin.  We initially assumed
expansion into an interstellar medium (ISM) with constant density
$n_0 \lsim 0.3 {\rm\ cm}^{-3}$, based on upper limits established by
Katsuda et al.  (2010), and adopted $E_{51} = 1$ based on the
spectral classification of Tycho's SN as a normal Type Ia event
(Krause et al. 2008).

For models that satisfied the above conditions, we varied additional
parameters that primarily impact the spectrum in subsequent \crhydro\
runs. These include the electron-to-proton number density ratio,
$\Kep$, a shape parameter for  spectral cutoff around the maximum
momentum, $\alpha_{cut}$ (see Lee et al. 2012), the ratio of the
wave damping and growth rates in the acceleration region, and the
spatial variation of the Alfv\'en speed in the shock precursor,
$f_{alf}$.

Using the shocked downstream and precursor magnetic field and plasma
density provided by the hydro calculations, we calculated the
synchrotron, IC, and nonthermal bremsstrahlung emission from the
relativistic electrons.  For the IC emission, we used seed photon
fields from the CMB, starlight, and IR emission from local dust,
to which we also added a local IR field from Tycho itself based on
{\sl Akari} observations \citep{IshiharaEtal2010}. Emission from
$\pi^0$-decay was calculated based on the model from \citet{Kamae06}.
Synchrotron, IC, and nonthermal bremsstrahlung emission from secondary
electrons was calculated as well, though these did not contribute
significantly to the overall emission.

For models that adequately reproduced the broadband SED for Tycho,
we used the predicted X-ray emission models as templates for fitting
\chandra\ spectra taken from regions along and just behind the
forward shock of the SNR (see Figure 1).

\subsection{Parameter Studies}

To arrive at a self-consistent model for Tycho's SNR, we first
explored the global parameter space by investigating the effects
of varying $d$, $n_0$, $B_0$, and $\xi.$ These results are summarized
in Figure 2 where we plot the variation with ambient density $n_0$
of the angular radius of Tycho, its angular expansion rate, and the
value of $\log E^2 dN/dE$ at fiducial energy values characterizing
the radio, nonthermal X-ray, high-energy and very-high-energy
$\gamma$-ray bands.\footnote{While Figure 2 displays results only
for $B = 5 \mu$G, models with values between $3 - 20 \mu$G were
investigated.} For each panel, horizontal bands indicate the
observed uncertainty range for the identified quantity, and the
different curves connect model calculations for different values
of the efficiency, distance, and unshocked magnetic field, as
indicated. 

The broad parameter space investigation above produced reasonable
models for input values of $d \sim 3$~kpc, $n_0 \sim 0.2 {\rm\
cm}^{-3}$, $B_0 \sim 5\ \mu$G, and $\xi \sim 3.7$. We then refined
these parameters along with $K_{e-p}$, and $f_{\rm alf}$ until the
predicted broadband spectrum provided a good agreement with the
observations.  The upper panel in Figure 3 shows the results for
our best model (Model A), whose parameters are summarized in Table
1.  We plot the radial variation of the density in order to best
illustrate the model positions for the FS, CD, and RS. The observed
average FS radius is indicated in green, and the uncertainty range
for the CD (RS) is indicated in magenta (red). While there is
actually an observed variation of about 6\% in the FS radius
for Tycho, all models were
constructed to yield the same outer shock position, to facilitate
more direct comparison. 

The dashed curve in the upper panel of Figure 3 corresponds to the
same input parameters as for Model A except that efficient particle
acceleration has been turned off. As expected, the FS/RS and FS/CD
radius ratios are smaller for Model A as a result of particle
acceleration. The RS position for Model A, which we note is sensitive
to the assumed density profile of the ejecta, is in good agreement
with that observed by Warren et al. (2005). The radius of the CD
falls considerably short of that inferred from the data. This may
be the result of Rayleigh-Taylor (R-T) instabilities at the CD
resulting in the penetration of ejecta into the shocked ISM. Such
instabilities in SN Ia remnants were studied by \citet{WC2001},
who found that these structures can extend as much as $\sim
11$\% beyond the CD radius at the dynamical age of Tycho's SNR.
This effective R-T extension region is indicated in cyan in the
upper panel of Figure 3.  Since the CD region identified by Warren
et al. (2005) corresponds to the interface between ejecta emission
and the shocked ISM, it is clear that ejecta-filled filaments
extending from the CD in Model A can explain the inferred position
of this interface. In addition, simulations show that, in the case
of efficient particle acceleration, such R-T structures can extend
to even larger distances \citep{BE2001, WB2013}.
Thus, we contend that the positions of the RS, FS, and CD
in Model A are in good agreement with observations.

\begin{figure}[t]
\epsscale{1.20}
\plotone{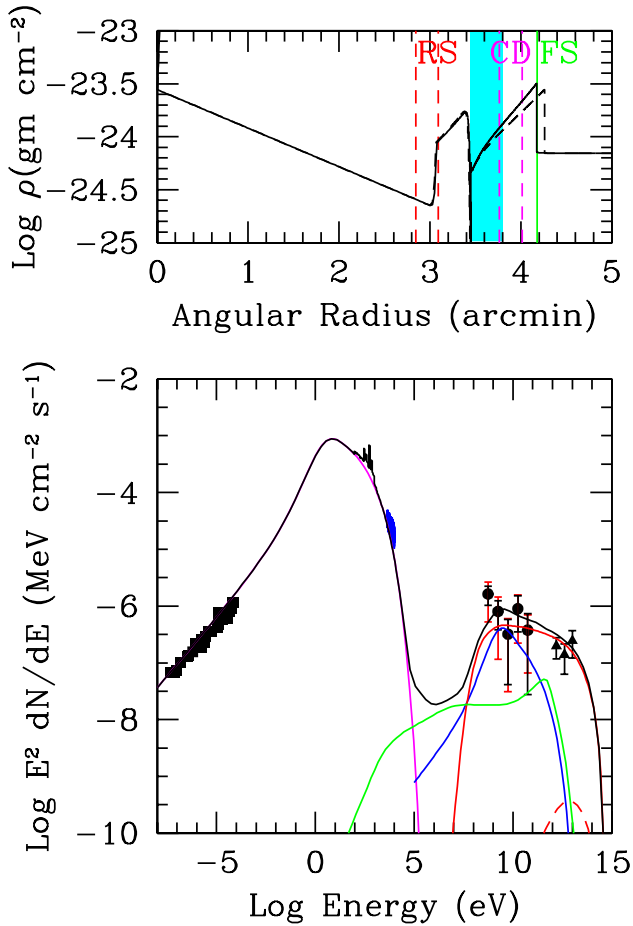}
\caption{
Top: Radial density profile for Model A. The black dashed curve
corresponds to the same parameters as Model A, but with the particle
acceleration effectively turned off. The green vertical line indicates
the position of the FS, while the magenta and red lines delineate
the CD and RS ranges reported by Warren et al. (2005). The cyan
band indicates the extent of expected R-T mixing based on the work
of Wang \& Chevalier 2001).  Bottom: Broadband SED for Model A,
compared with observed emission.  See text for description of
different curves.
\label{fig3}}
\end{figure}

\begin{table}[b]
\begin{center}
\caption{\footnotesize{\crhydro\ Model Parameters}}
\label{tab:specs}
\begin{tabular}{cccc}
\toprule
\noalign{\smallskip}
\noalign{\smallskip}
Parameter & Model A & Model B & Model C \\
\noalign{\smallskip}
\noalign{\smallskip}
\multicolumn{1}{c}{} & \multicolumn{3}{c}{\it Input} \\
\noalign{\smallskip}
$d \ (\rm kpc)$ & 3.18 & 3.40 & 2.90 \\
$n_0 \  (\rm{cm}^{-3})$ & 0.3 & 0.2 & 0.85 \\
$f_{\rm alf}$ & 0.7 & 0.5 & 0.5 \\
$\xi$ & 3.6 & 3.7 & 3.9 \\
$K_{ep}$ & 0.003 & 0.008 & 0.01 \\
\noalign{\smallskip}
\multicolumn{1}{c}{} & \multicolumn{3}{c}{\it Output} \\
\noalign{\smallskip}
$M_{sw}(M_\odot)$ & 2.47 & 2.02 & 5.33 \\
$R_{tot}$ & 4.64 & 4.48 & 4.23 \\
$E_{CR}/E{SN}$ & 0.16 & 0.11 & 0.07 \\
${\rm Eff}_{DSA}$ & 0.26 & 0.20 & 0.11 \\
$p_{max}/10^4 m_pc$ & 4.3 & 9.5 & 5.8 \\
\noalign{\smallskip}
\hline
\noalign{\smallskip}
\multicolumn{3}{l}{$E_{51} = 1$, $B = 5\ \mu$G, and $\alpha_{cut} = 0.5$ for all models.}
\end{tabular}
\end{center}
\end{table}

\subsection{Broadband SED}

The lower panel in Figure 3 presents the predicted spectrum for
Model A, along with the observed spectral measurements for Tycho.
Radio points were taken from \citet{RE92}, and the nonthermal X-ray
spectrum (5 - 10 keV) shown in blue is from {\sl Suzaku} observations
based on 444 ks of exposure (ObsIds: 500024010, 503085010, and
503085020), including 350 ks from the Tycho key project (PI: Hughes).
We use a local background from the outskirts of the field of view
and consider spectra only from the front-side illuminated chips
(XIS0 and XIS3), which we combined into a single, merged spectrum
and appropriate response file. The best fit over the 5-10 keV band
(using a model with a single power law and 5 Gaussians to describe
the thermal lines) yields a power-law photon index of $2.83 \pm
0.01$ with a normalization of $0.116 {\rm\ photons\ cm}^{-2}{\rm\
s}^{-1}{\rm\ keV}^{-1}$, values that are consistent with previous
measurements.  For Figure 3  we plot only the best fit power-law
component with estimated error bars.  Black circles represent data
from the Fermi LAT (Giordano et al. 2012) while triangles correspond
to data from VERITAS observations (Acciari et al. 2011). For the
models, the magenta curve represents the synchrotron emission, the
blue curve is from IC emission, the red curve is from $\pi^0$-decay,
and the green curve is the nonthermal bremsstrahlung emission. The
dashed red curve represents $\pi^0$-decay emission from escaping
protons, and the black curve is the sum of these model components.
Weak thermal line features in the soft X-ray band ($\sim 0.2 -
2.5$~keV), from the shocked ISM can be seen in the model. (We note
that interstellar absorption has not been applied to the SED)

The results of this model indicate that the $\gamma$-ray emission
from Tycho is dominated by $\pi^0$-decay at high energies, although
the IC emission at lower energies is very significant as well, and
makes a nearly equal contribution between 1 and 10 GeV.  This
conclusion is consistent with those presented in several recent
studies of Tycho, although our modeling approach contains important
differences, as we discuss below. We note that that the radio
spectrum presented in Figure 3 corresponds to the entire SNR 
while our modeled emission corresponds to the FS. While significant
particle acceleration is not expected at the RS, any such component
would not be accounted for in the model.

The results of the evolution for Tycho predicted by Model A are
summarized in Table 1. At its current age, the remnant has swept
up $\sim 2.5 M_\odot$ of ISM material. The remnant has placed roughly
16\% of the SNR kinetic energy into relativistic particles, with a
total DSA acceleration efficiency of $\sim 26$\%. The total compression
ratio from the forward shock, $R_{tot},$ is about 16\% higher than
expected for the case of no particle acceleration. The electron and
proton momentum spectra for Model A are shown in Figure 4; the
maximum proton energy is nearly 50~TeV.

\begin{figure}[t]
\epsscale{1.20}
\plotone{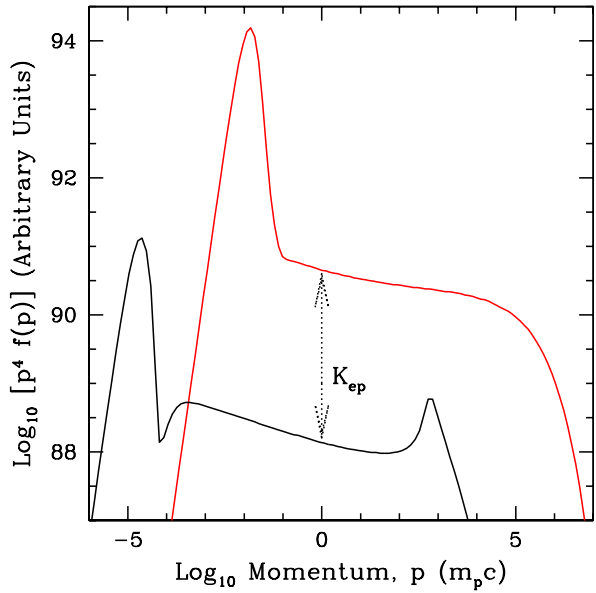}
\caption{
Proton (red) and electron (black) spectra $p^4 f(p)$ for Model A.
\label{fig4}}
\end{figure}

Figure 5 shows the results from our Model B (see Table 1), for which
the density is slightly lower than that of Model A. The dynamical
results are in agreement with the measurements of Warren et al.
(2005), although the RS position is just barely consistent with the
measured value. We note that recent {\sl Suzaku} measurements suggest
that the RS radius may be even smaller than previously recognized
\citep{YamaguchiEtal2014}, potentially indicating that the position
in this model (and perhaps that in Model A) is too large.
Model B represents our best model in which the $\gamma$-ray emission
is dominated by IC emission, although the highest energies are still
dominated by the $\pi^0$-decay component.
However, the overall fit to both the
GeV and TeV data is rather poor indicating that a model where 
leptons dominate the $\gamma$-ray flux is unlikely for Tycho.

In Figure 6 we present the time evolution of the FS (black) and RS
(red) angular speeds for our models. Horizontal lines indicate the
range of measured values for the angular expansion rate of the FS
and of the inner edge of the shocked ejecta which, particularly
for the FS, show considerable azimuthal variations (Katsuda et al.
2010). Both models A and B adequately reproduce the observed expansion 
rates.

\begin{figure}[t]
\epsscale{1.20}
\plotone{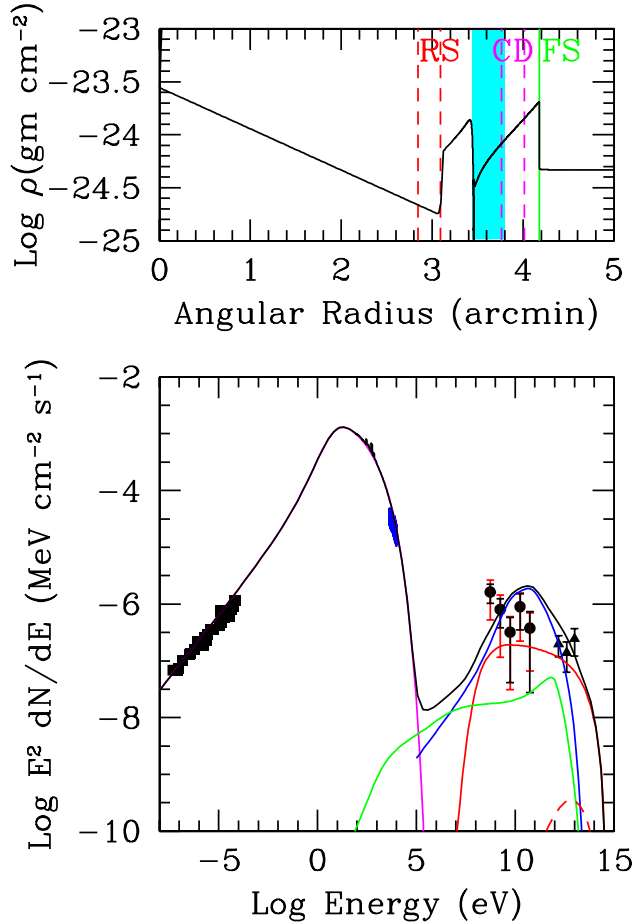}
\caption{
Same as Figure 5, for Model B.
\label{fig5}}
\end{figure}

\begin{figure}[t]
\epsscale{1.15}
\plotone{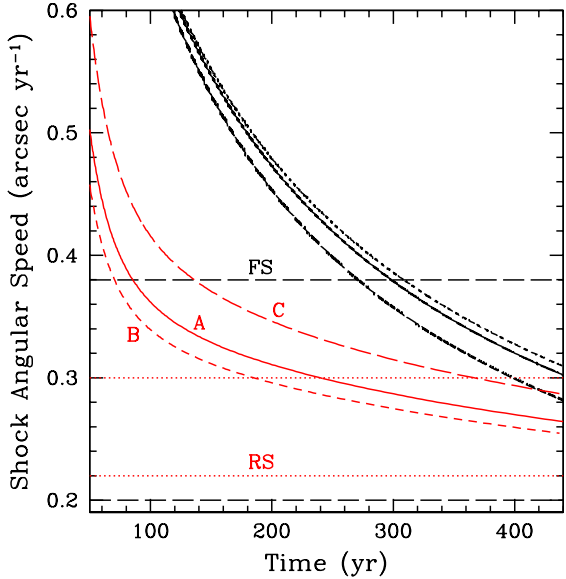}
\caption{
Time evolution of the FS and RS angular speeds for models described in
the text. Horizontal dashed curves outline the measured values for
the FS (black) and RS (red). Solid lines correspond to Model A while short
(long) dashed lines correspond to Model B (C).
\label{fig6}}
\end{figure}

\subsection{X-ray Emission}

The X-ray image in Figure 1 was created by merging observations
from a 750 ks \chandra\ observation carried out in 2009 (ObsIDs
10093-10097, 10902-10904, 10906). Standard cleaning procedures were
used on each individual data set, and the resulting images were
merged, resulting in a net exposure of 734.1~ks.  Spectra were
extracted from each of the rectangular regions in the northeastern,
northwestern, and western portions of the SNR indicated in Figure
1 (hereafter referred to as NE, NW, and W regions). These regions
were selected because the nonthermal emission is well separated
from the bright ejecta; they correspond closely with those used by
Cassam-Chena\"{\i} et al. (2007).  The large outer boxes in each
region were used for extraction of background spectra. For each
region, spectra from individual ObsIDs were combined using the {\sl
ciao} task {\sl dmtcalc}, and weighted arf (effective area)
and rmf (spectral redistribution) files were created
using the {\sl addresp} task.

The X-ray spectra from directly along the FS and from the region
immediately behind the shock in the NE, NW, and W regions of the
SNR are shown in Figures 7-9.  Fits were performed with {\sl xspec
version 12.7.1} using the thermal and nonthermal X-ray emission
models calculated in the \crhydro\ runs along with the interstellar
absorption model {\sl tbabs} \citep{WilmsEtal2000} with the abundance
set from \citet{angr89}.  The normalization of the two emission
components were tied together, but allowed to vary along with the
absorbing column density. We obtain good fits to the spectra from
regions 1 and 2 in Figure 1 using results from Model A (shown as a
histogram in each Figure), although emission from region 3 contains
a distinct excess of soft X-ray emission, discussed in Section 4.
Similar features were reported by Hwang et al. (2002). Given the
evidence for significant density variations around Tycho (Williams
et al. 2013), it is not surprising that some regions contain more
shocked ISM material than predicted by our 1-D models which average
over the entire SNR.  The fit parameters for regions 1 and 2 are
summarized in Table 2; region 3 was eliminated from these fits because
of clear ejecta contamination (see Section 4). The results from
fits to Model B (see Table 2) are similar to that from Model A; the
goodness-of-fit is actually slightly better than for Model A,
primarily due to a slightly flatter synchrotron component.

\begin{table*}
\begin{center}
\caption{\footnotesize{X-Ray Spectral Fits For Models A, B, and C}}
\label{tab:x-ray}
\begin{tabular}{c@{\hskip0.3in}ccc@{\hskip0.3in}ccc@{\hskip0.3in}ccc}
\toprule
\noalign{\smallskip}
\noalign{\smallskip}
\multicolumn{1}{c}{} & \multicolumn{3}{c}{NW$^a$} & \multicolumn{3}{c}{NE$^a$} & \multicolumn{3}{c}{W$^a$} \\ 
\noalign{\smallskip}
Parameter & A & B & C & A & B & C & A & B & C \\
\noalign{\smallskip}
$N_H^b$ & $6.3\pm0.02$ & $5.3\pm0.02$ & $8.9\pm0.01$ &  $6.6\pm0.02$ & $5.7\pm0.02$ & $8.7\pm0.02$ &  $8.0\pm0.02$ & $6.9\pm0.02$ & $8.9\pm0.01$  \\
$K_1^c$ & $71.8\pm1.6$ & $35.5\pm0.8$ & $80.8\pm1.7$ & $23.3\pm0.6$ & $11.4\pm0.4$ & $26.0\pm0.7$ & $101.0\pm1.3$ & $49.7\pm0.9$ & $110.0\pm1.7$ \\
$K_2^c$ & $18.4\pm0.6$ & $6.5\pm0.2$ & $13.5\pm0.4$ & $15.1\pm0.6$ & $5.1\pm0.2$ & $9.8\pm0.3$ & $50.7\pm1.2$ & $17.0\pm0.5$ & $33.9\pm0.8$ \\
$\chi_r^2$ & 1.3 & 1.2 & 1.5 & 1.1 & 1.0 & 1.2 & 1.5 & 1.3 & 1.6\\
dof & & 398 & & & 373 & & & 533\\
\noalign{\smallskip}
\noalign{\smallskip}
\noalign{\smallskip}
\hline
\noalign{\smallskip}
\multicolumn{10}{l}{a) Region sizes as fraction of total azimuth are 0.027 (NW), 0.023 (NE), and 0.032 (W).}\\
\multicolumn{10}{l}{b) Column density $\times 10^{21}{\rm\ cm}^{-2}$} \\
\multicolumn{10}{l}{c) Model normalization $\times 10^{-3}$}\\
\end{tabular}
\end{center}
\end{table*}

Because our model is spherically symmetric, the expected normalization
from the spectral fits is simply the fraction of the total azimuth
of the SNR covered by our extraction regions. For region 1, we find
that the normalization is in excellent agreement with the expected
geometric value in the NE region, but is a factor of 2.6 (3.1)
higher in the NW (W) region. This is consistent with typical
variations in the brightness around the rim of Tycho.  Moreover,
since we have selected brighter filaments in order to obtain good
spectra in small regions, it is not surprising that the results are
biased to normalizations somewhat higher than expected.  More notable
is the fact that the normalizations for region 2 falls below that
for region 1 in all three regions of the SNR. We find the same
behavior for region 3.

\begin{figure}[t]
\epsscale{1.15}
\plotone{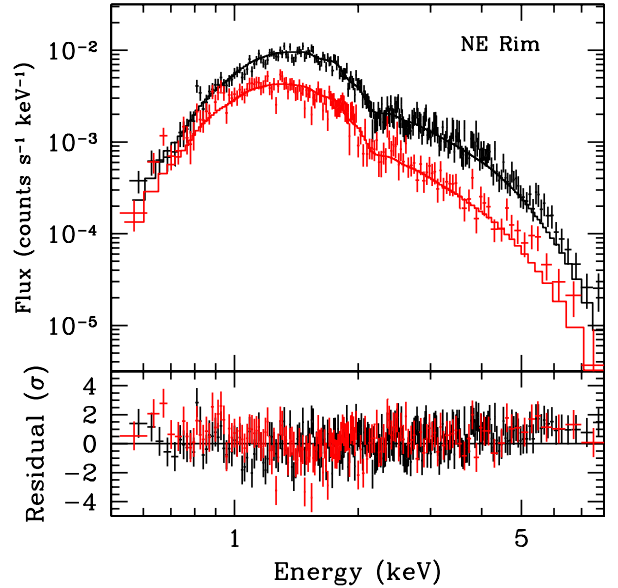}
\caption{
\chandra\ spectra from the NE rim of Tycho, compared with the
predictions from Model A. The upper (black) spectrum corresponds
to a region directly along the shock (region 1 in Figure 1) while
the lower (red) spectrum is taken from a region immediately behind
the shock (region 2).
\label{fig7}}
\end{figure}

The radial brightness profiles for the NE and W rims are shown in
Figure 10, where blue/red points correspond to X-ray data, and
green/magenta points represent radio profiles.  The accompanying
curves are the predicted brightness profiles from our Model A.  The
predicted X-ray profile is in reasonable good agreement with the
observations near the shock, but declines more slowly than the
observed brightness at larger distances behind the shock, consistent
with the relatively lower normalizations obtained in spectral fits
from these regions.  Meanwhile the predicted radio profile, while
correctly indicating a slow rise to a plateau-like region well
behind the shock, is less successful at reproducing the exact
profile. This behavior is similar to that found by Cassam-Chena\"{\i}
et al. (2007). Because the radio-emitting electrons do not suffer
significant radiative losses, the projected brightness profile
should show a continuous increase from the FS to the CD. The observed
climb to an early plateau may be an indication that these electrons
are confined to a shell that is considerably thinner than the region
between the FS and CD in our model. The physical mechanism
associated with such a picture is not obvious, although a reduction
in the FS/CD separation from R-T structures described above could
be partially responsible. Alternatively, the radial distribution
of the magnetic field in the postshock region may deviate from that
in our model. Cassam-Chena\"{\i} et al. (2007) suggest, for example,
that rapid damping of the field at the FS after its rapid initial
rise \citep{Pohl2005} could produce both the sharp rise and the
partial falloff of the radio emission, and that turbulent motions
from from the outermost ejecta could enhance the downstream field,
thus resulting in an increase in the radio synchrotron brightness
in this region.  The corresponding profiles from the NW region are
complicated by overlapping shocks that are evident in Figure 1, and
are thus not shown in Figure 10.

\begin{figure}[t]
\epsscale{1.15}
\plotone{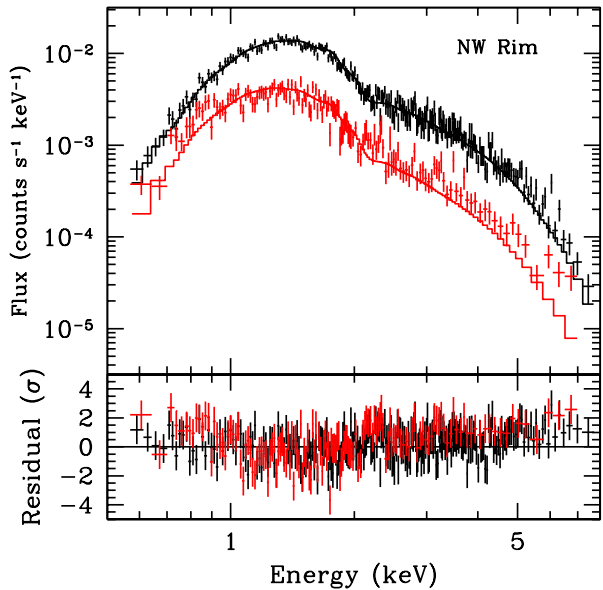}
\caption{
Same as Figure 8, for NW rim of Tycho.
\label{fig8}}
\end{figure}

\begin{figure}[t]
\epsscale{1.15}
\plotone{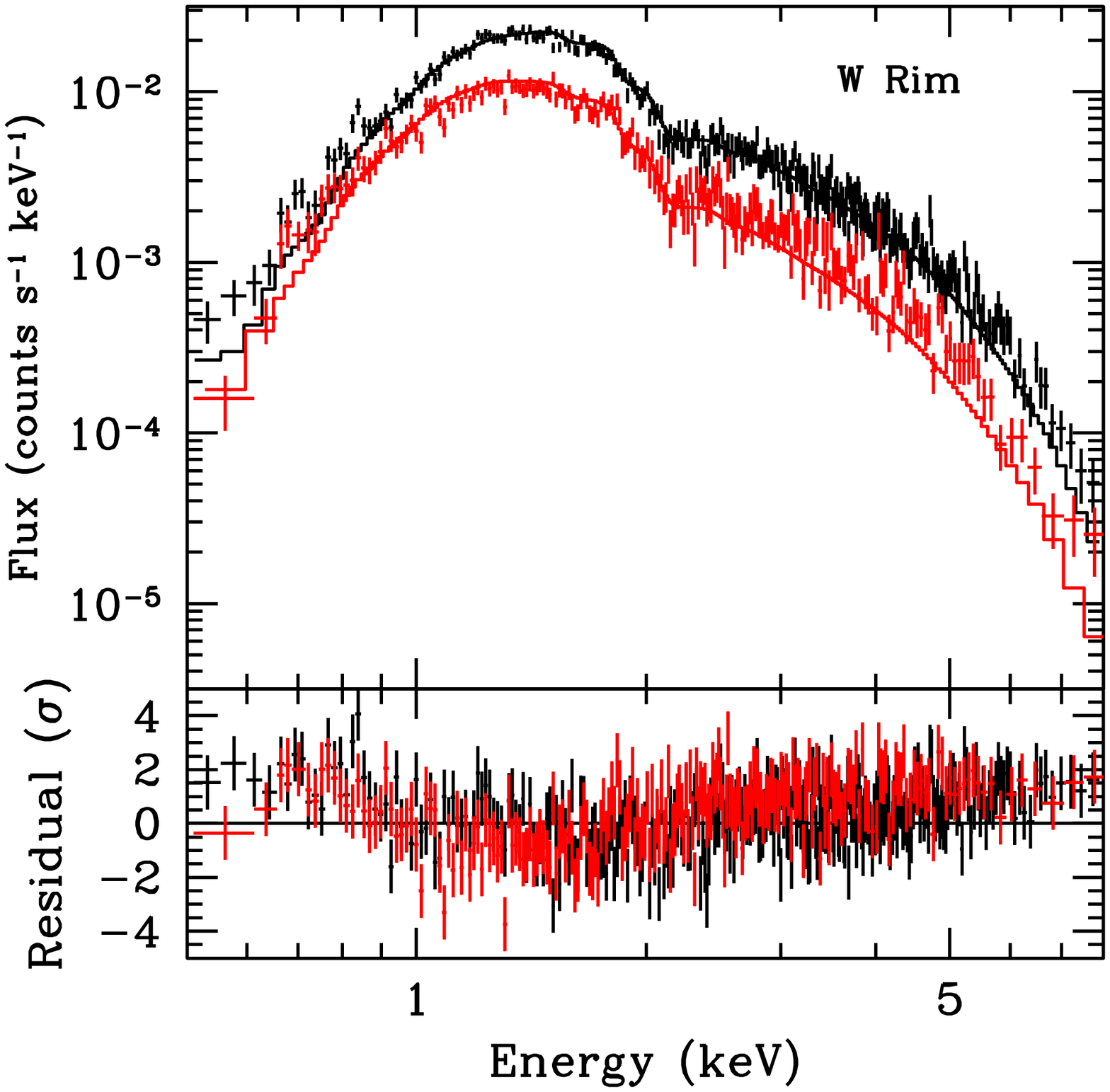}
\caption{
Same as Figure 8, for W rim of Tycho
\label{fig9}}
\end{figure}

\begin{figure}[t]
\epsscale{1.15}
\plotone{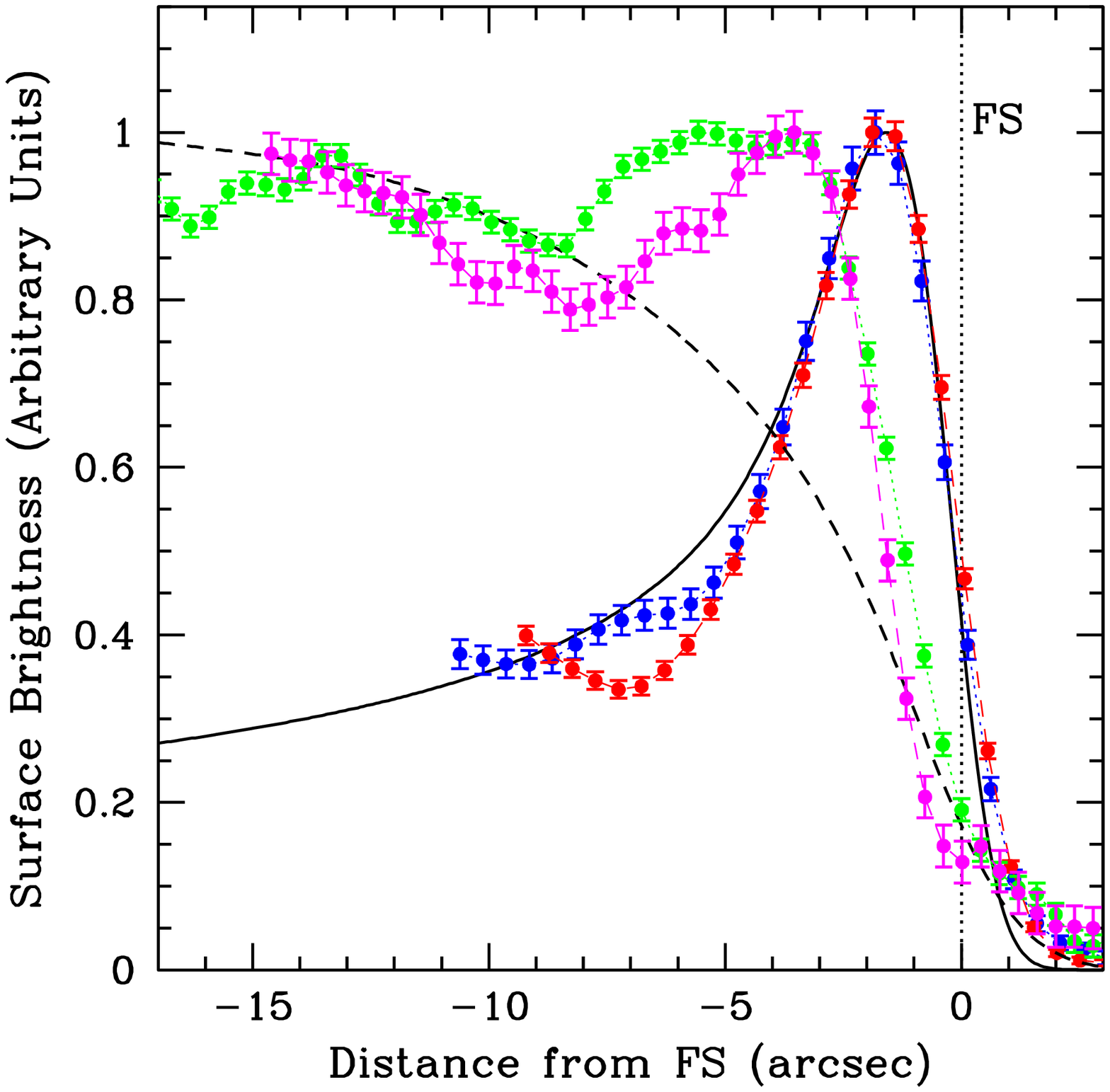}
\caption{
Observed X-ray and radio brightness profiles at the NE/W  rims of
Tycho, compared with predictions from Model A, described in the
text. Blue/red (green/magenta) points correspond to X-ray (radio)
data from the NE/W. Distances are measured from the position of the
FS. X-ray and radio data have been renormalized to produce a peak
brightness value of 1, and positions relative to the shock position
have been shifted to best align with data with the model profiles.
\label{fig10}}
\end{figure}

\subsection{Additional Models}

As noted in Section 1, models of the ejecta emission from Tycho by
Badenes et al. (2006) appear to require a somewhat higher density
than those derived from X-ray (Cassam-Chena\"{\i} et al. 2007,
Katsuda et al., 2010) and IR measurements (Williams et al. 2013).
Figure 11 presents our best results for a fixed ambient density of
$n_0 = 0.85 {\rm\ cm}^{-3}$ (our Model C), the lowest value reported
by Badenes et al. (2006).  The CD is located quite far behind the
position estimated by Warren et al. (2005), even if the effects of
R-T instabilities are considered, although it is important to note
that the exponential density profile assumed here differs from that
used by Badenes et al. (2006), which was based on actual explosion
models with stratified composition. The radio and $\gamma$-ray
components from the higher-density model also provide poor fits to
the data.  The integrated X-ray spectrum from Model C predicts
significant line emission, but the overall contribution of this
component to the small regions studied here is small; the
X-ray fits are formally worse than for Models A and B, but the
differences are not dramatic.  Thus, while we view the high density
model case to be very unlikely based on both the broadband emission
and the dynamical evolution of the shocks, it cannot be ruled out
by the X-ray data alone. The parameters for Model C are summarized
in Table 1.

Most recently \citet{ChiotellisEtal2013} have revisited the
investigation of the RS spectrum for Tycho for models that include
initial evolution in a small wind cavity. They find that the resulting
ionization characteristics provide a somewhat better match than
that obtained by Badenes et al. (2006) with a modest ambient density
of $\sim 0.4 {\rm\ cm}^{-3}$ and the presence of a wind-blown bubble
through which the remnant evolves at an early age. In this model,
the ambient density outside the cavity can be somewhat low, as
indicated by the X-ray and IR measurements, while the effective
density associated with the ionization can be somewhat higher due
to the initial expansion through the inner wind region. As this
model does not consider DSA, it cannot explain the observed
$\gamma$-ray emission.  Moreover, with DSA included, the proposed
density would result in a FS radius that is smaller than that
observed, as discussed in Section 1. We have thus investigated a
similar scenario, also considering the effects of efficient DSA on
the dynamical evolution.  Specifically, we inserted a relic
bubble with radius $0.4$~pc, blown by a stellar wind with $\dot{M}_w
= 3 \times 10^{-6} M_\odot {\rm\ yr}^{-1}$ and $v_w = 10 {\rm\ km\
s}^{-1}$, and allowed the SNR to evolve through the bubble and into
a uniform ambient medium. We find that using parameters 
similar to those from Model A, along with the inclusion of the
wind bubble, yields virtually no change in our results; a
slightly lower density is required ($n_0 = 0.28 {\rm\ cm}^{-3}$),
and the density profile shows structure from reflected shocks
associated with the wind bubble interaction, but the overall dynamics
are nearly identical and the broadband spectrum is indistinguisable
from that of Model A. This is to be expected, because the total
mass contained in the wind component is only $\sim 0.1 M_\odot$,
while the total swept-up mass in model A is $\sim 2.5 M_\odot$ (see
Table 1).  Thus, while an early effect on the ionization of the
ejecta (which we do not treat here) appears plausible, the impact
on the dynamical evolution is negligible.  Thus, we conclude that
the wind bubble model proposed by Chiotellis et al. (2013) can
produce results consistent with current dynamical and spectral
measurements of Tycho, though when the effects of cosmic ray
acceleration are taken into account, the required ambient density
is $\lsim 0.3 {\rm\ cm}^{-3}$.

\begin{figure}[t]
\epsscale{1.20}
\plotone{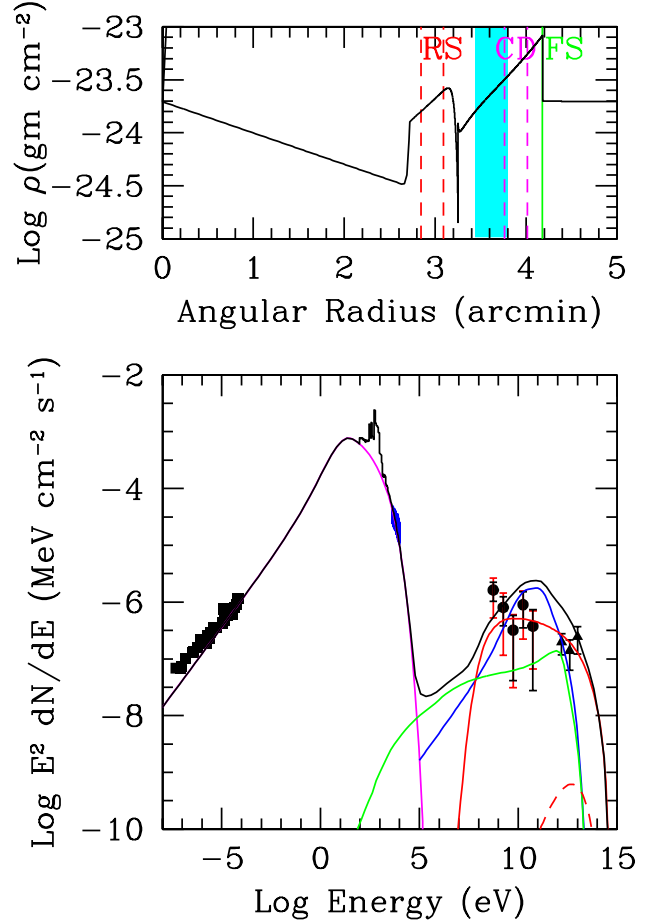}
\caption{
Same as Figure 3, but for Model C.
\label{fig11}}
\end{figure}

\section{Discussion}

Based upon our Model A, we conclude that Tycho's SNR has
evolved in a medium with an ambient density $n_0 \sim 0.3 {\rm\
cm}^{-3}$ and a magnetic field strength $B \sim 5\ \mu$G. The remnant
has undergone efficient acceleration of electrons and ions, with
$\sim 16$\% of the kinetic energy of the supernova explosion being
deposited into relativistic particles with $\Kep \sim 0.003$. The
DSA efficiency is $\sim 26$\% at the current epoch, the amplified
magnetic field in the immediate postshock region has a strength of
$\sim 180\ \mu$G, and roughly 11\% of the energy that has gone into
particle acceleration has been lost to escaping particles. The
remnant distance is $\sim 3.2$~kpc. This model is consistent with
the measured positions and expansion speeds of the FS and RS in
Tycho.

As noted in Section 3.1, with our models tuned to produce the
observed FS radius at the known age of Tycho, our Model A reproduces
the observed RS radius, but yields a CD position that falls short
of that estimated by Warren et al. (2005).  We suggested that R-T
filaments have resulted in ejecta material being mixed well beyond
the CD, toward the FS. In Figure 12 we show a spectrum from region
3 in the NE section of Tycho (see Figure 1) along with our best-fit
\crhydro\ model (black histogram). Significant residuals can be
seen at the positions of emission lines from Ne (0.92 keV), 
Si (1.85 keV), and S (2.45 keV), which
were subsequently fit by Gaussians with energies fixed at the line
energies for these elements. Similar results were reported by Hwang
et al. (2002) and Cassam-Chena\"{\i} et al.  (2007), and provide
strong evidence for the presence of ejecta mixed nearly all the way
to the FS, although the presence of significant amounts of
Ne is quite unexpected given the much lower abundance in
nucleosynthesis models for SN Ia. It must be noted that there
are numerous Fe-L features in the region around 1~keV, which
complicates the interpretation of this feature.

The broadband emission from Tycho's SNR has been modeled by a number
of researchers \citep[e.g.,][]{MC2012,AD2012,BKV2013,ZhangEtal2013}
with most concluding that the \gamray\ emission is dominated by
\pion. We reach a similar conclusion, although there are significant
differences between our model and previous ones. First, unlike all
other models, we simultaneously fit the broadband continuum and
X-ray line emission with a single \SC\ model.  As has been shown
previously \citep[e.g.,][]{EPSBG2007,PES2009,EPSR2010,LEN2012,
CSEP2012,LSENP2013}, the X-ray observations strongly constrain the
ambient density and can provide a distinction between \pion\ and
\IC\ emission at \gamray\ energies. Second, we are able to obtain
a satisfactory broadband fit without invoking an arbitrary
multi-component environment where some fraction of the swept-up
mass is in dense clumps as is done in \citet{BKV2013}.  Of course,
we do not argue that dense clumps do not exist; the morphology of
any SNRs is more complicated than the spherically symmetric model
we use.  At present, however, observational evidence of any such
clumpy environment surrounding Tycho is insufficient to provide
constrained parameters for such modeling.  From a modeling perspective,
there is thus little motivation to add a second component with an
additional set of unconstrained parameters when a one-component
model produces a satisfactory fit.

\begin{figure}[t]
\epsscale{1.20}
\plotone{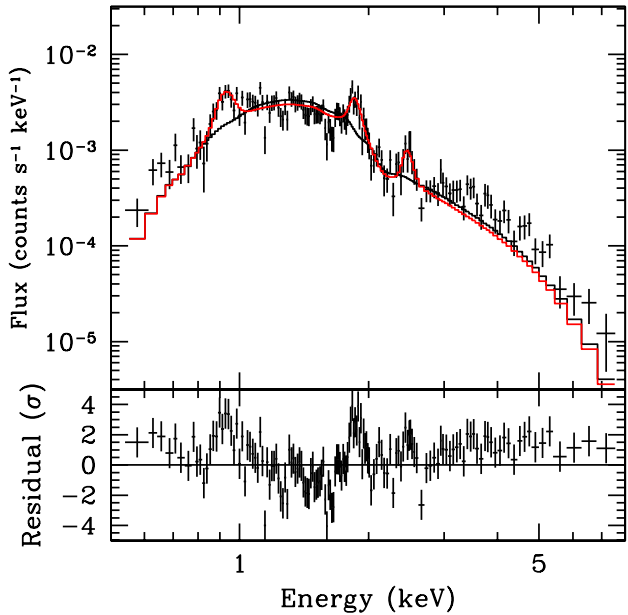}
\caption{
Spectrum from NE region 3 from Tycho (see Figure 1). The black histogram
corresponds to the best-fit absorbed \crhydro\ model for the corresponding 
projected region. Residual emission from Ne, Si, and S are evident,
indicating the presence of ejecta extended nearly all the way to the
FS. The red histogram corresponds to a \crhydro\ model with Gaussians
added at the energies expected for He-like emission from Ne, Si, and S,
although we note that significant Ne emission is not expected in 
spectra of Type Ia SNRs.
\label{fig12}}
\end{figure}

It has been argued (\citet{AD2012}) that a two-zone model in which
electrons accelerated at the FS shock emit in two distinct environments
provides a more realistic picture for Tycho's SNR than the many
one-zone models that have been considered. In this scenario, a thin
outer zone at the FS produces the X-ray \synch\ emission while,
inside the FS, electrons accumulate in a region of lower B-field.
The IC emission is produced from electrons in both zones, while the
synchrotron emission arises primarily from the outer zone.  Using
several unconstrained parameters to describe the zones, \citet{AD2012}
can fit the broadband continuum with the $\gamma$-ray emission
dominated by IC but they note that hadronic models are equally
viable.  As noted above, our evolving, continuous zone, spherically
symmetric model also has electrons (and ions) radiating in different
environments, shocked at different times, where the parameters are
determined from the hydro simulation coupled to the DSA calculation.

The most complete models of Tycho were presented by \citet{MC2012}
and \citet{BKV2013}. \citet{BKV2013} criticize \citet{MC2012} for
their presumably inconsistent treatment of diffusion.  Particle
diffusion is indeed critical for DSA; our treatment is described
in detail in \citet{LEN2012}. Briefly, the \crhydro\ simulation
includes a description of magnetic field amplification (MFA) in the
shock precursor, so the B-field turbulence is determined  as a
function of position in the precursor.  Our diffusion coefficient
is $D(x) = (vpc)/[3e \delta B(x)]$.  The strength of the field
variation $\delta B$ from MFA decreases with position in the precursor
as the free-escape boundary (FEB) is approached, resulting in an
increase in $D$ (i.e., scattering weakens) near the FEB.

Our model also includes non-adiabatic heating in the shock precursor.
The MFA produces turbulence, and the dissipation of turbulent energy
and heating of the background flow is parameterized.  The \alf\
wave speed, which impacts the effective compression ratio for DSA,
also varies with position in the precursor.  Importantly, there is
feedback between the shape and normalization of the accelerated
particle spectrum, the MFA and precursor heating, the speed of the
\alf\ scattering centers, the normalization of $D$ as a function
of position in the precursor, and $\pmax$.

We note that \crhydro\ also addresses the main concern \citet{BKV2013}
have concerning the \citet{MC2012} model, namely that the latter
authors assume a steep spectrum for the CR protons, in order to match
the observed spectrum at lower energies, but then assumed Bohm
diffusion for even the highest energy particles in order to obtain
emission of $\gamma$-rays beyond 400~MeV. This leads to a maximum
proton momentum $\pmax \sim 500$~TeV. Berezhko et al. (2013)
note that Bohm diffusion is inconsistent with such a steep spectrum,
and argue that a two-component spectrum is required. Here, we
do not make the same assumption. The spectrum of the accelerated
particles, the normalization of the diffusion coefficient, and
the resulting maximum momentum are all calculated self-consistently.
As noted above, we obtain $\pmax \sim 50$~TeV while still
reproducing the observed broadband spectrum.
In short, we believe the \crhydro\ model addresses
all of the criticisms \citet{BKV2013} make of the \citet{MC2012}
model, for which the results of our more detailed modeling  are in
good agreement.

In the results presented here, we do make the Bohm assumption, i.e.,
that the diffusion coefficient is equal to the particle gyroradius.
However, the gyroradius is determined with the local amplified
$\delta B(x)$ which varies with precursor position. At this point,
we do not believe that generalizing our fits to include a parameter
search with an arbitrary momentum dependence on the gyroradius is
warranted.

\section{Conclusions}

We have carried out a detailed study of the radial structure,
evolution, and broadband emission from Tycho's SNR. Using our
\crhydro\ code, we have investigated a range of parameters to
identify solutions that reproduce the observed dynamical properties
of the remnant and reproduce the observed broadband emission. A key
goal of our work was to assess the nature of the observed $\gamma$-ray
emission.  We find that, for our most successful model, this emission
is dominated by $\pi^0$-decay resulting from the collisions of
relativistic protons accelerated at the FS, with a significant
contribution at GeV energies arising from IC scatting of an ambient
photon field dominated by the CMB and local IR emission from Tycho
itself. The electron-to-proton ratio for injected particles is $\sim
0.003$.  Roughly 16\% of the SNR kinetic energy has been converted
into relativistic particles, and the impact of this on the SNR
evolution is a FS/RS radius ratio that is smaller than what would
result in a situation without particle acceleration.  This model
is also consistent with the measured positions and expansion speeds
of the FS and RS in Tycho, and predicts protons accelerated to
energies as high as $\sim 50$~TeV. The distance implied by this
model is $\sim 3.2$~kpc.

The projected brightness profiles from our 1-D model are in reasonable
agreement with those observed in the radio and X-ray bands at
discrete regions along the SNR rim. The predicted X-ray spectrum
from (projected) regions along the FS are in good agreement with
that observed by \chandra, while that from some regions immediately
behind the shock show evidence for density variations around Tycho
that are expected based on other studies (e.g., Williams et al.
2013).

The parameters we derive to explain the evolutionary state and history
for Tycho are in very good agreement with those from other works
that present models in which the $\gamma$-ray emission arises
predominantly from $\pi^0$-decay, particular those presented by
Morlino \& Caprioli (2012). Given our more complete treatment of
the evolution and emission, we conclude that these results are
robust and provide conclusive evidence for efficient acceleration
of cosmic-ray electrons and ions in Tycho's SNR.

\acknowledgments
The authors acknowledge important discussions with Brian Williams
and Carles Badenes concerning this work. P.S. acknowledges support
from NASA Contract NAS8-03060.  D.C.E. acknowledges support from
NASA grant NNX11AE03G. J.P.H. acknowledges support from Chandra
grant GO9-0078X and NASA {\sl Suzaku} grant NNX08AZ86G.




\bibliographystyle{aa} 
\bibliography{bib_POS}

\end{document}